\documentclass{article}

\usepackage{eusipco2011}                       

\usepackage{amsmath,graphicx}

\usepackage{psfrag}
\usepackage{epsfig}
\usepackage[latin1]{inputenc}
\usepackage{amsmath,amssymb,amsthm} 
\usepackage{amssymb}
\usepackage{url}      
\usepackage{color}
\usepackage{times}

\usepackage{graphicx}
\usepackage[english]{babel}

\title{Consistent Reconstruction of the Input\\ of an Oversampled Filter Bank from Noisy subbands}
\name{M. Abid$^1$, M. Kieffer$^{1,2}$, and B. Pesquet-Popescu$^1$}
\address{$^1$ Telecom ParisTech, Signal and Image Processing Department, \\
46 rue Barrault, 75634 Paris cedex 13, France\\
$^2$ on leave from L2S - CNRS - SUPELEC - Univ Paris-Sud, 91192 Gif-sur-Yvette, France}

\begin{document}

\maketitle

\begin{abstract}
This paper introduces a reconstruction approach for the input signal of an oversampled
filter bank (OFB) when the sub-bands generated at its output are quantized and transmitted 
over a noisy channel. This approach exploits the redundancy introduced by the OFB and the fact that the quantization noise is bounded. 

A maximum-likelihood estimate of the input signal is evaluated, which only considers the 
vectors of quantization indexes corresponding to subband signals that could have been 
generated by the OFB and that are compliant with the quantization errors.

When considering an OFB with an oversampling ratio of $3/2$ and a transmission of quantized subbands on an AWGN channel, compared to a classical decoder, 
the performance gains are up to $9$ dB in terms of SNR for the reconstructed signal, and $3$ dB in terms of channel SNR. 
\end{abstract}

\section{Introduction}

In classical communication systems based on Shannon separation principle \cite{Shannon48}, source coding and channel coding are optimized separately. However, due to delivery delay and processing complexity constraints, 
source and channel coding have to be performed on short to moderate-size vectors of source samples. When the channel conditions are better than those for which the channel code has been designed, some redundancy added by the channel coder is wasted. When they are worse than expected, transmission errors may not be efficiently corrected, and will have a detrimental effect on the reconstructed bitstream, see \cite{DuhamelKieffer09}.

Joint source and channel coding (JSCC) techniques have been considered to address these issues \cite{KatsaggelosBook07}. In this context, oversampled filter banks (OFB) \cite{Bolcskei97,Cvetkovic98} are particularly interesting, since they perform a signal decomposition into subbands, leaving some controlled redundancy among subbands. When transmitted over a communication channel, subbands are usually first quantized, introducing some background quantization noise in the transmitted subbands. Quantization indexes are then packetized and transmitted over a noisy channel. In absence of residual transmission errors, the redundancy introduced by the OFB in the subband domain has been shown to be helpful to combat quantization noise \cite{Goyal_IT98}. 

When channel impairments are badly corrected by channel decoders at receiver side, corrupted packets are obtained. Classical error detection techniques, such as CRCs or checksums, check the integrity of these packets  \cite{Kurose05}. When an erroneous packet is detected, retransmission may be asked, but in delay-constrained applications, this is not always possible. The content of the packet is then lost. The robustness of OFB and more generally of  frame expansions to the erasure of a whole subband has been evidenced, \emph{e.g.}, in \cite{Kovacevic02,RathSP04,PetrisorICASSP05}. The design freedom offered by OFB thanks to the introduced redundancy allows to construct synthesis filter banks that exploit the available samples at the receiver side to reconstruct the original signal with a minimal quadratic reconstruction error. These results have been extended in \cite{AkbariMMSP10} to the case of several subbands randomly affected by erasures, as is the case when quantized subbands are interleaved before being packetized. A progressive missing subband estimation technique has been developed in that case.    

When corrupted packets are not dropped, transmission errors result in corrupted quantization indexes, leading to subband samples corrupted by (large-variance) impulse noise. Samples not affected by transmission errors are also corrupted by (moderate-variance) quantization noise introduced at the transmitter side. A Gaussian-Bernoulli-Gaussian noise model \cite{Ghosh96} representing quite accurately the effect of quantization noise and transmission impairments has been used in \cite{LabeauChiang2004}. Parity-check filter banks associated to the analysis OFB have been exploited to build hypotheses tests determining whether a subband is affected by a transmission impairment at some time instant. These tests rely on computing a threshold whose value depends on the ratio of the variance of the impulse noise to that of the quantization noise (Impulse over quantization noise ratio, IQNR). The samples detected as corrupted are then corrected with a Bayesian estimator. An alternative approach to detect and correct corrupted subband samples has been proposed in \cite{Redinbo97,Redinbo2009} using Kalman filtering techniques. This method relies as well on a set of parameters to be chosen in advance (noise covariance matrices). 

The performance of all previously mentioned techniques is strongly dependent of the characteristics of the noise model. In practice, the quantization noise is not Gaussian, but more or less uniformly distributed, and the IQNR is not that high, leading to situations where the error detection and correction is difficult.  

The aim of this paper is to exploit the redundancy introduced by the OFB and to explicitly take into account the channel noise model and the bounded quantization noise. A suboptimal maximum-likelihood (ML) estimator is derived. The estimation is performed in the subspace of all \emph{consistent} indexes, \emph{i.e.}, indexes that can result from the quantization of a subband signal belonging to the subspace of subbands that may be generated at the output of the considered OFB. 

An implementation with a reasonable complexity of the proposed ML estimator is proposed. The main idea is to perform at each time instant an estimation of the vector of the most likely indexes with a sequential algorithm such as the \emph{M-algorithm} \cite{Anderson1991} and then eliminate those not deemed as \emph{consistent}. The consistency test is operated using interval analysis \cite{JaulinBook01}, but it could alternatively be done via the solution of several linear programs.

The rest of the paper is organized as follows. Section~\ref{sec:Transmission} describes
the considered transmission scheme based on an OFB. The formulation of the optimal ML estimator of the source samples from channel outputs is given in Section~\ref{sec:optimal}. A suboptimal estimator is presented in Section~\ref{sec:suboptimal} and the corresponding estimation algorithm is given in Section~\ref{sec:algo}. Preliminary simulation results are shown in Section~\ref{sec:res}, before providing some conclusions.

\begin{figure*}[htb]
\psfrag{b_i}[][][0.8]{$\mathbf{b}^{i}$}
\psfrag{r_i}[][][0.8]{$\mathbf{r}^{i}$}
\psfrag{E(z)}[][][1.2]{$E(z)$}
\psfrag{R(z)}[][][1.2]{$R(z)$}
\psfrag{Q}[][][1.2]{$Q$}
\psfrag{Q^(-1)}[][][1.2]{$Q^{-1}$} 
\psfrag{z^(-1)}[][][0.8]{$z^{-1}$}
\psfrag{x_n}[][][0.8]{$\mathbf{x}_{n}$} 
\psfrag{hx_n}[][][0.8]{$\widehat{\mathbf{x}}_{n}$}
\psfrag{x_Ni}[][][0.8]{${x}_{Ni}$} 
\psfrag{x_Ni+1}[][][0.8]{${x}_{Ni+1}$}
\psfrag{x_Ni+N-1}[][][0.8]{${x}_{Ni+N-1}$} 
\psfrag{x_i}[][][0.8]{$\mathbf{x}_{i}$}
\psfrag{hx_Ni}[][][0.8]{$\widehat{x}_{Ni}$} 
\psfrag{hx_Ni+1}[][][0.8]{$\widehat{x}_{Ni+1}$}
\psfrag{hx_Ni+N-1}[][][0.8]{$\widehat{x}_{Ni+N-1}$} 
\psfrag{y_Mi+M-1}[][][0.8]{${y}_{Mi+M-1}$}
\psfrag{y_Mi+1}[][][0.8]{${y}_{Mi+1}$} 
\psfrag{y_Mi}[][][0.8]{${y}_{Mi}$}
\psfrag{u_Mi+M-1}[][][0.8]{${u}_{Mi+M-1}$} 
\psfrag{u_Mi+1}[][][0.8]{${u}_{Mi+1}$}
\psfrag{u_Mi}[][][0.8]{${u}_{Mi}$} 
\psfrag{hu_Mi+M-1}[][][0.8]{$\tilde{u}_{Mi+M-1}$}
\psfrag{hu_Mi+1}[][][0.8]{$\tilde{u}_{Mi+1}$} 
\psfrag{hu_Mi}[][][0.8]{$\tilde{u}_{Mi}$}
\psfrag{hy_Mi+M-1}[][][0.8]{$\tilde{y}_{Mi+M-1}$} 
\psfrag{hy_Mi+1}[][][0.8]{$\tilde{y}_{Mi+1}$}
\psfrag{hy_Mi}[][][0.8]{$\tilde{y}_{Mi}$} 
\centering \includegraphics[width=10cm]{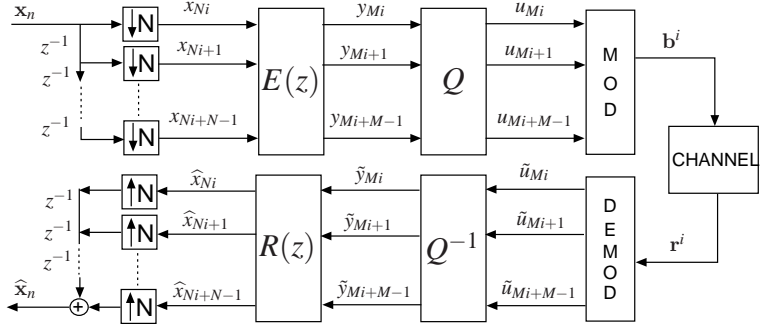}
\caption{Transmission scheme based on an $M$-band oversampled filter bank\vspace{3cm}}
\label{schema_BFS}
\end{figure*}
\section{Coding and transmission scheme}
\label{sec:Transmission}

Figure~\ref{schema_BFS} describes a typical transmission scheme
based on an $M-$band OFB with a downsampling factor of $N\leq M$,
introducing a redundancy in the subbands of $M/N$. The analysis filterbank
consists of $M$ FIR analysis filters $\left\{ \mathbf{h}_{m}\right\} _{m=0}^{M-1}$
with maximal length $N\times(L+1)$. The corresponding polyphase representation
of these filters is a $M\times N$ matrix $E(z)$. 
At each instant $i$, the vector $\mathbf{x}^{i}=\left(x_{Ni},\dots,x_{Ni+N-1}\right)^{T}$
is placed at the input of the OFB and the vector $\mathbf{y}^{i}=\left(y_{Mi},\dots,y_{Mi+M-1}\right)^{T}$
is obtained at its output. The relation in the temporal domain between
the input and the output of the OFB is then \begin{equation}
\mathbf{y}^{i}=\sum_{l=0}^{L}\mathbf{E}_{l}\mathbf{x}^{i-l}=\mathbf{E}_{L:0}\mathbf{x}^{i-L:i},\label{TempDomain_Relation}\end{equation}
where
$\mathbf{x}^{i-L:i}=\left(\left(\mathbf{x}^{i-L}\right)^{T},\dots,\left(\mathbf{x}^{i}\right)^{T}\right)^{T}$
contains all input samples affecting the OFB output at time $i$
and $\mathbf{E}_{L:0}=\left(\mathbf{E}_{L},\dots,\mathbf{E}_{0}\right)$
is a $M\times(L+1)N$ matrix formed by a sequence of $M\times N$ matrices
$\mathbf{E}_{l},\mbox{ }l=0,\dots,L$
that can be constructed from $\left\{ \mathbf{h}_{m}\right\} _{m=0}^{M-1}$
\cite{Vaidyanathan93}.
Since $E\left(z\right)$ represents a FIR filter bank, one can find
a $\left(M-N\right)\times M$ polyphase matrix $P\left(z\right)$
such that $P\left(z\right)E\left(z\right)=0 \:\:\:\: \forall z\in\mathbb{C}$ and that represents a FIR \emph{parity-check} filter bank, see \emph{Proposition 1} in
\cite{LabeauChiang2004}.
One can then write
\begin{equation}
P\left(z\right)=\sum_{l=0}^{L'}\mathbf{P}_{l}z^{-l}.\label{eq:Parity}
\end{equation}
Since $P\left(z\right)E\left(z\right)=0$, one has
\begin{equation}
\sum_{l=0}^{L'}\mathbf{P}_{l}\mathbf{y}^{i-l}=\mathbf{P}_{L':0}\mathbf{y}^{i-L':i}=\mathbf{0},\label{eq:Paritychecktemp}
\end{equation}
where
$\mathbf{y}^{i-L':i}=\left(\left(\mathbf{y}^{i-L'}\right)^{T},\dots,\left(\mathbf{y}^{i}\right)^{T}\right)^{T}$.
This property allows to determine whether a subband signal may be obtained at the output of
an OFB.

For the transmission, each component $y_{Mi+m}$, $m=0,\dots,M-1$,
of the vector $\mathbf{y}^{i}$ is quantized using a scalar quantizer
with a step-size $\Delta_{m}$. The resulting quantization indexes
$u_{Mi+m}$ are binarized to get a sequence $\mathbf{b}_{Mi+m}=\mathbf{b}(u_{Mi+m})$ of $R_{m}$ bits. The whole vector $\mathbf{u}^{i}=\left(u_{Mi},\dots,u_{Mi+M-1}\right)^{T}$
of quantized indexes is then represented by a binary sequence
$\mathbf{b}^{i}=\left(\mathbf{b}^{T}_{Mi},\dots,\mathbf{b}^{T}_{Mi+M-1}\right)$
of $\sum_{m=0}^{M-1}R_{m}$ bits that are BPSK modulated
and then transmitted over a memoryless channel
with a transition probability $g\left(r|b\right)$. A vector $\mathbf{r}^{i} =\left(\mathbf{r}^{T}_{Mi},\dots,\mathbf{r}^{T}_{Mi+M-1}\right)^{T}$
of binary-, real-~, or complex-valued samples is finally obtained at
channel output. 

A classical decoder would perform a hard decision on $\mathbf{r}^{i}$ to get some estimate $\widetilde{\mathbf{u}}^{i}$ of $\mathbf{u}^{i}$. After inverse quantization of $\widetilde{\mathbf{u}}^{i}$,
the received subbands $\widetilde{\mathbf{y}}^{i}$ are obtained. Finally, the reconstruction is performed using the pseudo-inverse of the $E(z)$, whose polyphase representation $R(z)=(E(z)^{T}E(z))^{-1}E(z)^{T}$ is a $N\times M$ matrix such that $R(z)E(z) = I_{N\times N}$, where $I_{N\times N}$ is the $N\times N$ identity matrix. Nevertheless, this estimator may produce estimated subbands $\widetilde{\mathbf{y}}^{i}$ which may not be produced by the considered OFB. The proposed estimator addresses this issue.

\section{Optimal ML estimator}
\label{sec:optimal}

At receiver side the ML estimate of the input vector at time $i$
assuming that all channel outputs have been gathered in a vector $\mathbf{r}$
is \begin{equation}
\widehat{\mathbf{x}}^{i}=\arg\max_{\widetilde{\mathbf{x}}^{i}\in[\mathbf{x}]}p(\mathbf{r}|\widetilde{\mathbf{x}}^{i}),\label{Eq:ML_Est}
\end{equation}
where $\left[\mathbf{x}\right]$ is a vector of intervals (or \emph{box})
to which all the vectors $\mathbf{x}^{i}$ are known to belong. The
box $\left[\mathbf{x}\right]$ may be obtained from the dynamic of the input signal. It is aussmed to be known \emph{a priori}
by the receiver. Since the channel is memoryless and the maximal length
of the impulse response of the analysis filters is $N\times(L+1)$ one gets
\begin{equation}
\widehat{\mathbf{x}}^{i}=\arg\max_{\widetilde{\mathbf{x}}^{i}\in[\mathbf{x}]}p(\mathbf{r}^{i:i+L}|\widetilde{\mathbf{x}}^{i})\label{eq:ML_est_2}\end{equation}
where $\mathbf{r}^{i:i+L}=\left(\left(\mathbf{r}^{i}\right)^{T},\dots,\left(\mathbf{r}^{i+L}\right)^{T}\right)^{T}$.
Let $\mathcal{U}$ be the set of all vectors $\widetilde{\mathbf{u}}^{i}$
of indexes that may be obtained at the output of the quantizers. This
set, containing at most $\prod_{m=0}^{M-1}2^{R_{m}}$ elements, is independent
of $i$ since the characteristics of the scalar quantizers do not
depend on time. Then, the conditional probability in \eqref{eq:ML_est_2} becomes 
\begin{equation}
\begin{split}p\left(\mathbf{r}^{i:i+L}|\widetilde{\mathbf{x}}^{i}\right) & =\sum_{\widetilde{\mathbf{u}}^{i:i+L}\in\mathcal{U}^{L+1}}p\left(\mathbf{r}^{i:i+L},\widetilde{\mathbf{u}}^{i:i+L}|\widetilde{\mathbf{x}}^{i}\right),\\
 & =\sum_{\widetilde{\mathbf{u}}^{i:i+L}\in\mathcal{U}^{L+1}}p\left(\mathbf{r}^{i:i+L}|\widetilde{\mathbf{u}}^{i:i+L},\widetilde{\mathbf{x}}^{i}\right)p\left(\widetilde{\mathbf{u}}^{i:i+L}|\widetilde{\mathbf{x}}^{i}\right)
\end{split}
\label{eq:ML_est_3}
\end{equation}
The channel output $\mathbf{r}^{i:i+L}$ depends only on the channel
input $\widetilde{\mathbf{u}}^{i:i+L}$. Hence,
$\widetilde{\mathbf{x}}^{i}$ does not provide any additional knowledge
on $\mathbf{r}^{i:i+L}$ once $\widetilde{\mathbf{u}}^{i:i+L}$ is known, $i.e.$, 
$\widetilde{\mathbf{x}}^{i} \longleftrightarrow \widetilde{\mathbf{u}}^{i:i+L} \longleftrightarrow \mathbf{r}^{i:i+L}
$ forms a Markov chain.
Then \eqref{eq:ML_est_3} becomes
\begin{equation}
p\left(\mathbf{r}^{i:i+L}|\widetilde{\mathbf{x}}^{i}\right)  =\sum_{\widetilde{\mathbf{u}}^{i:i+L}\in\mathcal{U}^{L+1}}p\left(\mathbf{r}^{i:i+L}|\widetilde{\mathbf{u}}^{i:i+L}\right)p\left(\widetilde{\mathbf{u}}^{i:i+L}|\widetilde{\mathbf{x}}^{i}\right)
\label{eq:ML_est_4}
\end{equation}
Using the fact that the channel is memoryless, the first term $p\left(\mathbf{r}^{i:i+L}|\widetilde{\mathbf{u}}^{i:i+L}\right)$
of \eqref{eq:ML_est_4} is easily obtained from the channel transition probability
\begin{equation}
\begin{split}
p\left(\mathbf{r}^{i:i+L}|\widetilde{\mathbf{u}}^{i:i+L}\right)&
 = \prod_{\ell = 0}^{L} p(\mathbf{r}^{i+\ell}|\widetilde{\mathbf{u}}^{i+\ell})\\
&
 = \prod_{\ell = 0}^{L} \hspace{0.3cm}\prod_{m = 0}^{M-1} p(\mathbf{r}_{M(i+\ell)+m}|\widetilde{u}_{M(i+\ell)+m}), \\
&
 = \prod_{\ell = 0}^{L} \hspace{0.3cm}\prod_{m = 0}^{M-1} g(\mathbf{r}_{M(i+\ell)+m}|\mathbf{b}(\widetilde{u}_{M(i+\ell)+m})).
\end{split}
\label{eq::mod_eq}
\end{equation}
The term $g(\mathbf{r}_{M(i+\ell)+m}|\mathbf{b}(\widetilde{u}_{M(i+\ell)+m}))$ of \eqref{eq::mod_eq}
is then obtained as the product of $R_{m}$ channel transition probabilities corresponding to the
$R_m$ bits in $\mathbf{b}(\widetilde{u}_{M(i+\ell)+m})$.

The second term $p\left(\widetilde{\mathbf{u}}^{i:i+L}|\widetilde{\mathbf{x}}^{i}\right)$
of \eqref{eq:ML_est_4}
is much more complex to evaluate. Moreover, the number of terms of
the sum in \eqref{eq:ML_est_4} is in general prohibitively large.
A suboptimal estimator is thus introduced in the next section.


\section{Suboptimal ML estimator}

\label{sec:suboptimal}

A suboptimal ML estimator for $\mathbf{x}^{i}$ is obtained when considering
only the channel output at time $i$. Thus, one gets
\begin{equation}
\hat{\mathbf{x}}^{i}=\arg\max_{\widetilde{\mathbf{x}}^{i}\in[\mathbf{x}]}p(\mathbf{r}^{i}|\widetilde{\mathbf{x}}^{i})\label{Eq:ML_Est_SubOpt}
\end{equation}
and 
\begin{equation}
\begin{split}p\left(\mathbf{r}^{i}|\widetilde{\mathbf{x}}^{i}\right) & =\sum_{\widetilde{\mathbf{u}}^{i}\in\mathcal{U}}p\left(\mathbf{r}^{i}|\widetilde{\mathbf{u}}^{i}\right)p\left(\widetilde{\mathbf{u}}^{i}|\widetilde{\mathbf{x}}^{i}\right)\end{split}
\label{eq:ML_estimation_SubOpt1}
\end{equation}

To evaluate $p\left(\widetilde{\mathbf{u}}^{i}|\widetilde{\mathbf{x}}^{i}\right)$,
one knows that a vector of quantized indexes $\widetilde{\mathbf{u}}^{i}$
is produced when the value $\mathbf{y}^{i}$ taken by the random vector
$\mathbf{Y}^{i}$ of OFB outputs belongs to some box $\left[\mathbf{y}^{i}\left(\widetilde{\mathbf{u}}^{i}\right)\right]=\left[\mathbf{y}^{i}\left(\widetilde{\mathbf{u}}^{i}\right)-\frac{\mathbf{\Delta}}{2},\mathbf{y}^{i}\left(\widetilde{\mathbf{u}}^{i}\right)+\frac{\mathbf{\Delta}}{2}\right]$,
where $\mathbf{y}^{i}\left(\widetilde{\mathbf{u}}^{i}\right)$ is obtained
from inverse quantization of $\widetilde{\mathbf{u}}^{i}$ and $\mathbf{\Delta}=\left(\Delta_{1},\dots,\Delta_{M}\right)^{T}$.
Then \begin{equation}
p\left(\mathbf{r}^{i}|\mathbf{X}^{i}=\widetilde{\mathbf{x}}^{i}\right)=\sum_{\widetilde{\mathbf{u}}^{i}\in\mathcal{U}}p\left(\mathbf{r}^{i}|\widetilde{\mathbf{u}}^{i}\right)p\left(\mathbf{Y}^{i}\in\left[\mathbf{y}^{i}\left(\widetilde{\mathbf{u}}^{i}\right)\right]|\mathbf{X}^{i}=\widetilde{\mathbf{x}}^{i}\right),\label{eq:ML_est_Sub_2}\end{equation}
 where $\mathbf{X}^{i}$ is the random vector at the input of the
analysis OFB. One may show that the second term of \eqref{eq:ML_est_Sub_2}
may be written as \begin{eqnarray*}
p\left(\mathbf{Y}^{i}\in\left[\mathbf{y}^{i}\left(\widetilde{\mathbf{u}}^{i}\right)\right]|\widetilde{\mathbf{x}}^{i}\right)\\
& \hspace{-6.5cm}= & \hspace{-3.5cm}\int\limits_{\left[\mathbf{x}\right]^{L}}
p\left(\mathbf{Y}^{i}\in\left[\mathbf{y}^{i}\left(\widetilde{\mathbf{u}}^{i}\right)\right],\widetilde{\mathbf{x}}^{i-L:i-1}|\widetilde{\mathbf{x}}^{i}\right) d\widetilde{\mathbf{x}}^{i-L:i-1}\\
 & \hspace{-6.5cm}= & \hspace{-3.5cm}\int\limits_{ \left[\mathbf{x}\right]^{L}} p\left(\mathbf{E}_{L:0}\mathbf{X}^{i-L:i}\in\left[\mathbf{y}^{i}\left(\widetilde{\mathbf{u}}^{i}\right)\right]|\widetilde{\mathbf{x}}^{i-L:i}\right)p\left(\widetilde{\mathbf{x}}^{i-L:i-1}|\widetilde{\mathbf{x}}^{i}\right) d\widetilde{\mathbf{x}}^{i-L:i-1} \end{eqnarray*}
 Then, \eqref{Eq:ML_Est_SubOpt} becomes \[
\hat{\mathbf{x}}^{i}=\arg\max_{\widetilde{\mathbf{x}}^{i}\in[\mathbf{x}]}\sum_{\widetilde{\mathbf{u}}^{i}\in\mathcal{U}}
\hspace{-0.15cm}
p\left(\mathbf{r}^{i}|\widetilde{\mathbf{u}}^{i}\right)
\hspace{-0.2cm}
\int\limits_{\left[\mathbf{x}\right]^{L}} \hspace{-0.2cm}
f\left(\widetilde{\mathbf{u}}^{i},\widetilde{\mathbf{x}}^{i-L:i}\right)p\left(\widetilde{\mathbf{x}}^{i-L:i-1}|\widetilde{\mathbf{x}}^{i}\right) d\widetilde{\mathbf{x}}^{i-L:i-1}
\hspace{-0.3cm},\]
 where
$f\left(\widetilde{\mathbf{u}}^{i},\widetilde{\mathbf{x}}^{i-L:i}\right)=p\left(\mathbf{E}_{L:0}\mathbf{X}^{i-L:i}\in\left[\mathbf{y}^{i}\left(\widetilde{\mathbf{u}}^{i}\right)\right]|\widetilde{\mathbf{x}}^{i-L:i}\right)$.
For a specific value $\widetilde{\mathbf{u}}^{i}$ one has \[
f\left(\widetilde{\mathbf{u}}^{i},\widetilde{\mathbf{x}}^{i-L:i}\right)=\begin{cases}
1 & \mbox{if \ensuremath{\mathbf{E}_{L:0}\mathbf{\widetilde{x}}^{i-L:i}\in\left[\mathbf{y}^{i}\left(\widetilde{\mathbf{u}}^{i}\right)\right]}}\\
0 & \mbox{otherwise.}\end{cases}\]
 The term $f\left(\widetilde{\mathbf{u}}^{i},\widetilde{\mathbf{x}}^{i-L:i}\right)$
accounts for the fact that all values in $\mathcal{U}$ cannot be obtained
at the quantized output of the OFB. One then obtains \begin{equation}
\hat{\mathbf{x}}^{i}=\arg\max_{\widetilde{\mathbf{x}}^{i}\in[\mathbf{x}]}\sum_{\widetilde{\mathbf{u}}^{i}\in\mathcal{U}}p\left(\mathbf{r}^{i}|\widetilde{\mathbf{u}}^{i}\right)
\hspace{-0.3cm}
\int\limits_{\mathcal{S}\left(\widetilde{\mathbf{u}}^{i},\widetilde{\mathbf{x}}^{i}\right)}\hspace{-0.3cm}
p\left(\widetilde{\mathbf{x}}^{i-L:i-1}|\widetilde{\mathbf{x}}^{i}\right) d\widetilde{\mathbf{x}}^{i-L:i-1},\label{Eq:ML_Est_SubOpt1}\end{equation}
where
\begin{equation*}
\mathcal{S}\left(\widetilde{\mathbf{u}}^{i},\widetilde{\mathbf{x}}^{i}\right) = 
\left\{
\widetilde{\mathbf{x}}^{i-L:i-1}\in\left[\mathbf{x}\right]^{L} |
\mathbf{E}_{L:0}\mathbf{\widetilde{x}}^{i-L:i}\in\left[\mathbf{y}^{i}\left(\widetilde{\mathbf{u}}^{i}\right)\right]
\right\}
\end{equation*}

Assume further that the estimation process in the previous instants $j=i-L,\dots,i-1$
has been able to provide boxes $\left[\widehat{\mathbf{x}}^{j}\right]$
such that $\mathbf{x}^{j}\in\left[\widehat{\mathbf{x}}^{j}\right]$.
Then, \eqref{Eq:ML_Est_SubOpt1} becomes \begin{equation}
\begin{split}\hat{\mathbf{x}}^{i} & =\arg\max_{\widetilde{\mathbf{x}}^{i}\in[\mathbf{x}]}\sum_{\widetilde{\mathbf{u}}^{i}\in\mathcal{U}}p\left(\mathbf{r}^{i}|\widetilde{\mathbf{u}}^{i}\right)\hspace{-1cm}
\int\limits_{\left[\widehat{\mathbf{x}}^{i-L:i-1}\right] \cap \mathcal{S}\left(\widetilde{\mathbf{u}}^{i},\widetilde{\mathbf{x}}^{i}\right)
}
\hspace{-1cm}p\left(\widetilde{\mathbf{x}}^{i-L:i-1}|\widetilde{\mathbf{x}}^{i}\right) d\widetilde{\mathbf{x}}^{i-L:i-1}\end{split}
\label{Eq:ML_Est_SubOpt2}\end{equation}
Consider now for each $\widetilde{\mathbf{u}}^{i}\in\mathcal{U}$ the
set \begin{eqnarray*}
\mathcal{X}^{i}\left(\widetilde{\mathbf{u}}^{i}\right) & = & \left\{ \widetilde{\mathbf{x}}^{i}\in[\mathbf{x}]|
\left[\widehat{\mathbf{x}}^{i-L:i-1}\right] \cap \mathcal{S}\left(\widetilde{\mathbf{u}}^{i},\widetilde{\mathbf{x}}^{i}\right) \neq \emptyset
\right\}\\
 & = & \left\{ \widetilde{\mathbf{x}}^{i}\in[\mathbf{x}]|\exists\widetilde{\mathbf{x}}^{i-L:i-1}\in\left[\widehat{\mathbf{x}}^{i-L:i-1}\right]\right.\\
 &  & \left.\mbox{ such that }E_{0}\mathbf{\widetilde{x}}^{i}\in\left[\mathbf{y}^{i}\left(\widetilde{\mathbf{u}}^{i}\right)\right]-\mathbf{E}_{L:1}\widetilde{\mathbf{x}}^{i-L:i-1}\right\}
\end{eqnarray*}
This set contains all values of the input vector at time $i$ for
which there exists some value of the preceding input vectors that
leads to quantized OFB output indexes represented by $\widetilde{\mathbf{u}}^{i}$.
$\mathcal{X}^{i}\left(\widetilde{\mathbf{u}}^{i}\right)$ is either empty
or is a polytope \cite{Walter97}. The set $\mathcal{U}$ can be then
partitioned into two subsets $\mathcal{U}_{0}=\left\{ \widetilde{\mathbf{u}}^{i}\in\mathcal{U}|\mathcal{X}^{i}\left(\widetilde{\mathbf{u}}^{i}\right)=\emptyset\right\} $
and $\mathcal{U}_{1}=\left\{ \widetilde{\mathbf{u}}^{i}\in\mathcal{U}|\mathcal{X}^{i}\left(\widetilde{\mathbf{u}}^{i}\right)\neq\emptyset\right\} $.
Therefore \eqref{Eq:ML_Est_SubOpt2} becomes \begin{equation}
\hat{\mathbf{x}}^{i}=\arg\max_{\widetilde{\mathbf{x}}^{i}\in[\mathbf{x}]}\sum_{\widetilde{\mathbf{u}}^{i}\in\mathcal{U}_{1}}p\left(\mathbf{r}^{i}|\widetilde{\mathbf{u}}^{i}\right)\hspace{-1cm}\int\limits_{\left[\widehat{\mathbf{x}}^{i-L:i-1}\right] \cap \mathcal{S}\left(\widetilde{\mathbf{u}}^{i},\widetilde{\mathbf{x}}^{i}\right)
}
\hspace{-1cm}p\left(\widetilde{\mathbf{x}}^{i-L:i-1}|\widetilde{\mathbf{x}}^{i}\right)d\widetilde{\mathbf{x}}^{i-L:i-1}.
\label{Eq:ML_Est_SubOpt3}\end{equation}

\subsection{Other approximations}

In \eqref{Eq:ML_Est_SubOpt3}, instead of considering all terms of
the sum over all $\widetilde{\mathbf{u}}^{i}\in\mathcal{U}_{1}$, we only
consider the term corresponding to \begin{equation}
\hat{\mathbf{u}}^{i}=\arg\max_{\widetilde{\mathbf{u}}^{i}\in\mathcal{U}_{1}}p\left(\mathbf{r}^{i}|\widetilde{\mathbf{u}}^{i}\right).\label{Eq:ML_Est_u}
\end{equation}
This is the ML estimate of the quantized indexes at time $i$ accounting
for the fact that $\hat{\mathbf{u}}^{i}\in\mathcal{U}_{1}$,\emph{
i.e.}, that it can be obtained as a quantized output of the considered
OFB. The classical ML estimate would consider a maximization over
$\mathcal{U}$ in \eqref{Eq:ML_Est_u}.
Using \eqref{Eq:ML_Est_u}, \eqref{Eq:ML_Est_SubOpt3} becomes then
\begin{equation}
\hat{\mathbf{x}}^{i}=\arg\max_{\widetilde{\mathbf{x}}^{i}\in[\mathbf{x}]}
\int\limits_{\left[\widehat{\mathbf{x}}^{i-L:i-1}\right] \cap \mathcal{S}\left(\hat{\mathbf{u}}^{i},\widetilde{\mathbf{x}}^{i}\right)
}
\hspace{-1cm}
p\left(\widetilde{\mathbf{x}}^{i-L:i-1}|\widetilde{\mathbf{x}}^{i}\right)d\widetilde{\mathbf{x}}^{i-L:i-1}.
\label{Eq:ML_Est_SubOpt4}\end{equation}
where the integral in \eqref{Eq:ML_Est_SubOpt4} vanishes when $\widetilde{\mathbf{x}}^{i}\notin\mathcal{X}^{i}\left(\mathbf{\hat{u}}^{i}\right)$.

Evaluating the function to maximize in \eqref{Eq:ML_Est_SubOpt4} is
still complicated. Thus, one evaluates the set of values $\mathcal{X}^{i}\left(\mathbf{\hat{u}}^{i}\right)$
on which the interval does not vanish, or more precisely an outer-approximation
of this set. Considering some $\mathbf{\widetilde{u}}^{i}\in\mathcal{U}$,
an \emph{outer} approximation $\left[\mathbf{x}^{i}\left(\mathbf{\widetilde{u}}^{i}\right)\right]$
of $\mathcal{X}^{i}\left(\mathbf{\widetilde{u}}^{i}\right)$ may be evaluated in
several ways. One may consider solving several linear programming
problems, or using tools from interval analysis \cite{JaulinBook01}.
If an empty outer-approximation is obtained for some $\widetilde{\mathbf{u}}^{i}$,
then $\widetilde{\mathbf{u}}^{i}\notin\mathcal{U}_{1}$. This allows to
construct an iterative algorithm that evaluates at each $i$ the most
likely vectors of indexes and keeps only the most likely one which also
belongs to $\mathcal{U}_{1}$.

\subsection{Parity-check test (PCT)}

To determine whether some vector of indexes belongs to $\mathcal{U}_{1},$
one may alternatively use the parity-check polyphase matrix $P\left(z\right)$.
One knows that if $\mathbf{y}^{i-L':i}$ contains only vectors corresponding
to the output of the OFB with polyphase matrix $E\left(z\right)$,
then \eqref{eq:Paritychecktemp} is necessarily satisfied. Consider
now a box $\left[\mathbf{y}^{i-L':i}\right]$. If $
\mathbf{0}\notin\mathbf{P}_{L':0}\left[\mathbf{y}^{i-L':i}\right],
$
then, there cannot be any sequence of $L'$ vectors $\mathbf{y}^{i-L':i}\in\left[\mathbf{y}^{i-L':i}\right]$
such that $\mathbf{y}^{i-L':i}$ is the output of the OFB with polyphase
matrix $E\left(z\right)$. This allows to build a quick test to determine
whether $\left[\mathbf{y}^{i}\left(\widetilde{\mathbf{u}}^{i}\right)\right]$
may contain an OFB output at time $i$ and consequently whether $\widetilde{\mathbf{u}}^{i}$
has to be further considered. For that purpose, a box $\left[\widehat{\mathbf{y}}^{i-L':i-1}\right]$
containing previously verified outputs has to be available. Then,
for some $\widetilde{\mathbf{u}}^{i}$, if\begin{equation}
\mathbf{0}\notin\sum_{l=1}^{L'}\mathbf{P}_{l}\left[\mathbf{\widehat{y}}^{i-l}\right]+\mathbf{P}_{0}\left[\mathbf{y}^{i}\left(\widetilde{\mathbf{u}}^{i}\right)\right]\label{eq:TestSynd2}\end{equation}
then  $\widetilde{\mathbf{u}}^{i}$ does not belong to $\mathcal{U}_{1}$.

The proposed algorithm is described in Section~\ref{sec:algo}. 
\section{Estimation algorithm}
\label{sec:algo}

This section describes the OFB input signal estimation algorithm using
the signal measured at channel output. 
\begin{enumerate}
\item \textbf{Input:} $\left[\mathbf{\widehat{x}}^{i-L:i-1}\right]$, $\left[\mathbf{\widehat{y}}^{i-L':i-1}\right]$,
\textbf{Output: }$\left[\mathbf{\widehat{x}}^{i}\right]$, $\left[\mathbf{\widehat{y}}^{i}\right]$. 
\item Initilization: $k=1$ ; $\ell=1$ ; $\mathcal{L}_{cand}^{\mathcal{U}_{1}}=\emptyset$. 
\item Sort the vectors of indexes $\widetilde{\mathbf{u}}^{i}$ in the decreasing
likelihood $p\left(\mathbf{r}^{i}|\widetilde{\mathbf{u}}^{i}\right)$.
Keep the $N_{\max}$ most likely candidates $\mathcal{L}_{cand}=\left\{ \widetilde{\mathbf{u}}^{i}\left(1\right),\dots,\widetilde{\mathbf{u}}^{i}\left(N_{\max}\right)\right\} $. 
\item Do

\begin{enumerate}
\item Evaluate $\left[\mathbf{x}^{i}\left(\widetilde{\mathbf{u}}^{i}\left(k\right)\right)\right]\subset[\mathbf{x}]$
satisfying $\mathbf{E}_{0}\left[\mathbf{x}^{i}\left(\widetilde{\mathbf{u}}^{i}\left(k\right)\right)\right]\subset\left[\mathbf{y}^{i}\left(\widetilde{\mathbf{u}}^{i}\left(k\right)\right)\right]-\mathbf{E}_{L:1}\left[\mathbf{\widehat{x}}^{i-L:i-1}\right]$. 
\item If $\left[\mathbf{x}^{i}\left(\widetilde{\mathbf{u}}^{i}\left(k\right)\right)\right]\neq\emptyset$
then $\mathcal{L}_{cand}^{\mathcal{U}_{1}}=\left\{ \mathcal{L}_{cand}^{\mathcal{U}_{1}},\widetilde{\mathbf{u}}^{i}\left(k\right)\right\} $ 
\item $k=k+1$; 
\end{enumerate}
\item while $k\leq N_{\max}$; 
\item If $\mathcal{L}_{cand}^{\mathcal{U}_{1}}=\emptyset$ then $\left[\mathbf{\widehat{x}}^{i}\right]=\left[\mathbf{x}^{i}\left(\widetilde{\mathbf{u}}^{i}\left(1\right)\right)\right]$
; $\widehat{\mathbf{u}}^{i}=\widetilde{\mathbf{u}}^{i}\left(1\right)$;
Indicate an error. End.\\
 Else $N_{\max}^{\mathcal{U}_{1}}=\left|\mathcal{L}_{cand}^{\mathcal{U}_{1}}\right|$; 
\item Do
\item Apply the parity-check test on $\left[\mathbf{y}^{i}
\left(
\mathcal{L}_{cand}^{\mathcal{U}_{1}}\left(l\right)
\right)
\right]$ 

\begin{enumerate}
\item If $\mathbf{0}\in\sum_{l=1}^{L'}\mathbf{P}_{l}\left[\mathbf{\widehat{y}}^{i-l}\right]+\mathbf{P}_{0}\left[\mathbf{y}^{i}
\left(
\mathcal{L}_{cand}^{\mathcal{U}_{1}}\left(l\right)
\right)
\right]$,
then $\left[\mathbf{\widehat{x}}^{i}\right]=\left[\mathbf{x}^{i}
\left(
\mathcal{L}_{cand}^{\mathcal{U}_{1}}\left(l\right)
\right)
\right]$,
$\left[\mathbf{\widehat{y}}^{i}\right]=\left[\mathbf{y}^{i}
\left(
\mathcal{L}_{cand}^{\mathcal{U}_{1}}\left(l\right)
\right)
\right],$
$\widehat{\mathbf{u}}^{i}=
\mathcal{L}_{cand}^{\mathcal{U}_{1}}\left(l\right)
$; End.
\item Else $l=l+1$; 
\end{enumerate}
\item while $l\leq N_{\max}^{\mathcal{U}_{1}}$.
\item $\left[\mathbf{\widehat{x}}^{i}\right]=
\left[\mathbf{x}^{i}
\left(
\mathcal{L}_{cand}^{\mathcal{U}_{1}}\left(1\right)
\right)\right]$,
$\left[\mathbf{\widehat{y}}^{i}\right]=
\left[\mathbf{y}^{i}
\left(
\mathcal{L}_{cand}^{\mathcal{U}_{1}}\left(1\right)
\right)\right],\widehat{\mathbf{u}}^{i}=
\mathcal{L}_{cand}^{\mathcal{U}_{1}}\left(1\right)
$,
Indicate an error. End.
\end{enumerate}

In the first call of the algorithm ($i=1$), the vectors $\left[\mathbf{\widehat{x}}^{j}\right]$$,\mbox{ \ensuremath{j=i-L,\dots,i-1}}$
and $\left[\mathbf{\widehat{y}}^{j}\right]$, $j=i-L',\dots$,$i-1$
are initialized to $\left[\mathbf{0},\mathbf{0}\right]$. 

Step 3.
can be performed with an M-algorithm \cite{Anderson1991}. 

Step 4a. may be performed using linear programming or interval analysis \cite{JaulinBook01}.
In the latter case, the problem is to enclose the $n-$ dimensional polytope
$\mathcal{X}^{i}\left(\mathbf{\widetilde{u}}^{i}\right)$ by an $n-$ dimensional box $\left[\mathbf{x}^{i}\left(\mathbf{\widetilde{u}}^{i}\right)\right]$.
The box $\left[\mathbf{x}^{i}\left(\mathbf{\widetilde{u}}^{i}\right)\right]$
is an outer approximation of $\mathcal{X}^{i}\left(\mathbf{\widetilde{u}}^{i}\right)$.

At step 6., $\mathcal{L}_{cand}^{\mathcal{U}_{1}}$ contains
the most likely vectors of indexes $\widetilde{\mathbf{u}}^{i}$ for
which one was not able to prove that they are not in $\mathcal{U}_{1}.$
The PCT is then applied to eliminate some candidates
in $\mathcal{L}_{cand}^{\mathcal{U}_{1}}$. 

The most likely vector of indices $\widehat{\mathbf{u}}^{i}\in\mathcal{U}_{1}$
is then obtained in the output of the algorithm as well as 
$\left[\mathbf{x}^{i}\left(\mathbf{\widehat{u}}^{i}\right)\right]$
and a box $\left[\mathbf{\widehat{y}}^{i}\right]=\left[\mathbf{y}^{i}\left(\widehat{\mathbf{u}}^{i}\right)\right]$
containing the noise-free OFB output at time $i$.

Since the \emph{a priori} distribution of the input $\mathbf{X}^{i-L:i}$
is not known in general, the estimate $\widehat{\mathbf{x}}^{i}$ in \eqref{Eq:ML_Est_SubOpt4}
could be chosen at the center of the box $\left[\mathbf{x}^{i}\left(\mathbf{\widehat{u}}^{i}\right)\right]$
which represents an easily evaluated approximation of the centroid of the polytope $\mathcal{X}^{i}\left(\mathbf{\widehat{u}}^{i}\right)$.
Besides $\left[\mathbf{\widehat{x}}^{i}\right]=\left[\mathbf{x}^{i}\left(\mathbf{\widehat{u}}^{i}\right)\right]$
may be used for the estimation of $\widehat{\mathbf{x}}^{i+1}$.

When no satisfying solution can be found, at Step 6. or at Step 10.,
the estimate corresponding to the classical ML estimate of $\mathbf{u}^{i}$ is
chosen, even if this may have a detrimental impact on the next estimates.
An error message is thus provided.



\section{Experimental results}

\label{sec:res}

This section provides preliminary results obtained using the algorithm
of Section \ref{sec:algo}. Two types of one-dimensional signals are considered:
a discrete-valued signal consisting of Lines~$55$ to $58$ of Lena.pgm
and a discrete-time correlated Gaussian signal with a correlation ratio of $0.9$. 
For each signal, $2000$ samples have been considered. 

The resulting signals go through
an OBF based on Haar filters, with $M=6$ subbands and a downsampling
rate $N=4$. Rate allocation is performed to equalize the variance
of the quantization noise in each subband. A BPSK modulation of the
binarized indexes is performed before transmission over
an AWGN channel with SNR from $6$~dB to $11$~dB. The results have
been averaged over $250$ noise realizations for both signals.
\begin{figure}[htbp] 
\centering \includegraphics[width=\columnwidth]{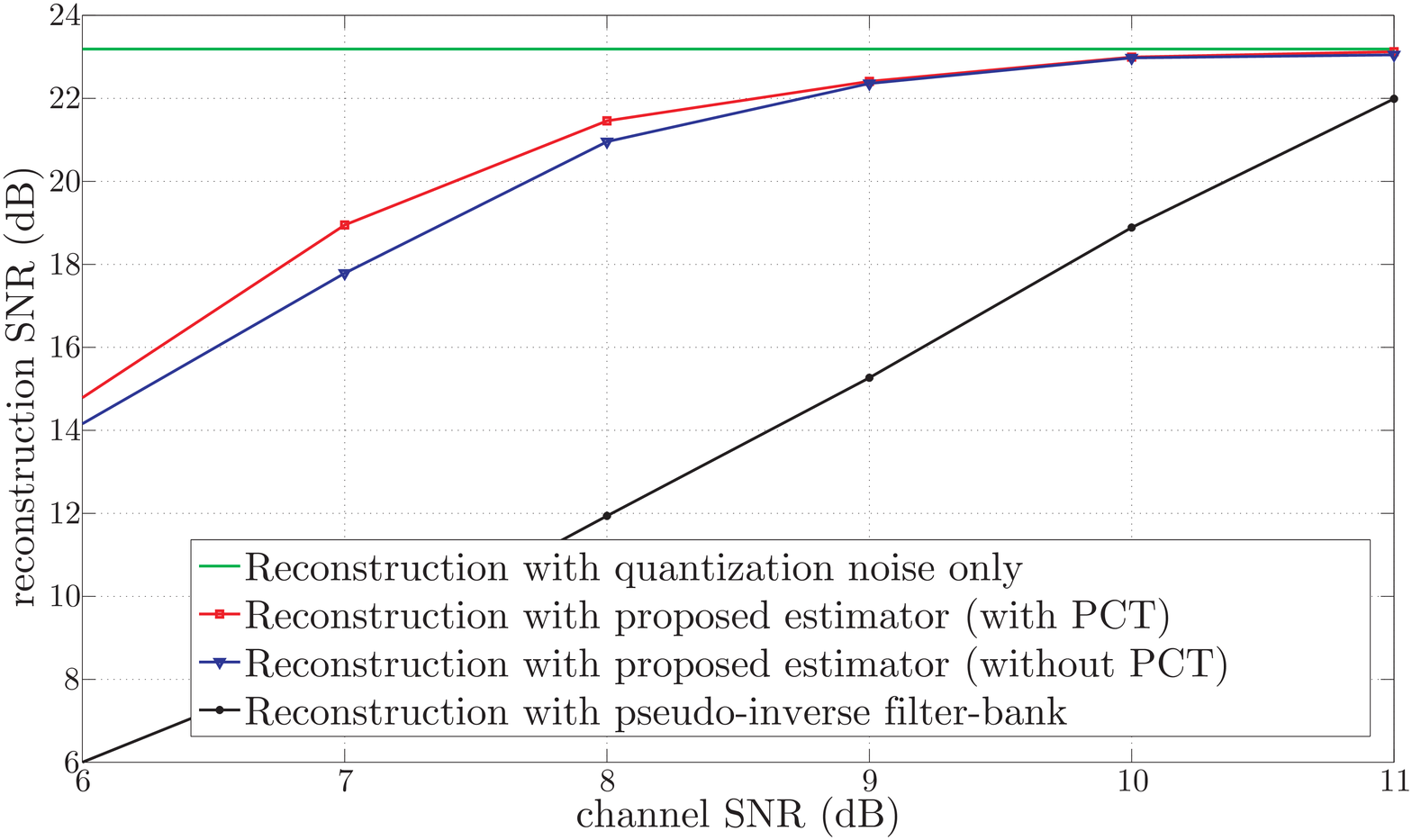}
\caption{SNR of reconstructed signals as a function of channel SNR for Lines~$55$ to $58$ of Lena.pgm}
\label{courbe_res_Lena}
\end{figure} 
\begin{figure}[htbp] 
\centering \includegraphics[width=\columnwidth]{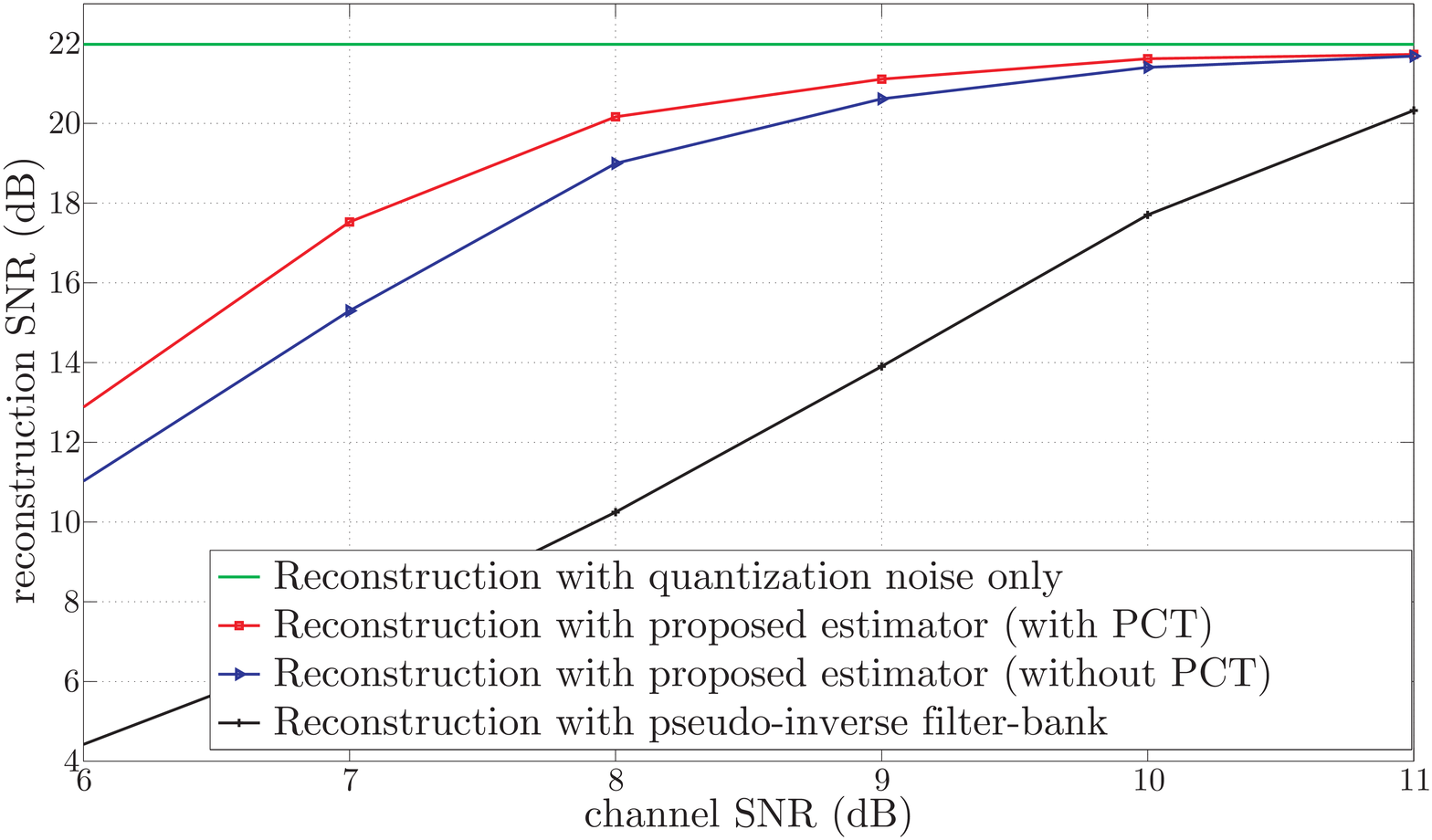}
\caption{SNR of reconstructed signals as a function of channel SNR for correlated Gaussian signal}
\label{courbe_res_gaussian}
\end{figure} 

Figures~\ref{courbe_res_Lena} and~\ref{courbe_res_gaussian}~show the average reconstruction SNR as
a function of the channel SNR. Three reconstruction methods are
compared. The OFB input signal is reconstructed with the proposed
estimator (with and without PCT) and with the conventional reconstruction method that uses the synthesis filter $R(z)$ on the most likely vector of indexes, after its inverse quantization,  at each time
instant, without paying any attention to the fact that it may not
belong to $\mathcal{U}_{1}$. The number of candidates considered at each iteration has been
set to $N_{\max}=20$. The reference reconstruction SNR corresponds
to a signal corrupted only by quantization noise.

For both signals and without the use of the PCT, a gain
of about $8$~dB in reconstruction SNR is observed for a channel SNR of $7$~dB and a gain of
about $2.5$~dB in the channel SNR is observed for a signal SNR of $14$~dB . The use
of the PCT in the proposed algorithm improves the performance in terms
of reconstructed signal SNR. For a channel SNR level of $7$~dB, the gain in signal SNR is about $1$~dB for the first signal and of $2$~dB for the
second one.

For the Gaussian correlated signal, the correlation between samples has not been explicitly taken into account here.
It may be taken into account in \eqref{Eq:ML_Est_SubOpt4} to further improve estimation performance.
\section{Conclusion}

In this paper we presented an estimation method that exploits the redundancy
introduced by OFB as well as the fact that the noise introduced by the quantization operation is bounded. 

The obtained experimental results have shown the improvement brought by this approach when compared to a classical estimation technique. The use of a parity-check filter bank allows to increase the performances of the method.
This motivates us to propose methods that use the PCT in a more efficient way. 
Future work will be dedicated as well to extending the proposed approach to images and video. 



\end{document}